\documentclass[11pt]{article}

\usepackage[normalem]{ulem}
\usepackage[swedish,english]{babel}

\usepackage{color}
\usepackage{xcolor}
\usepackage{amssymb}
\usepackage{amsmath}

\usepackage{ae}
\usepackage{slashed}
\usepackage{graphics}
\usepackage{graphicx}
\usepackage{epstopdf}
\usepackage{url}
\usepackage{hyperref}

\usepackage{cite}

\usepackage{times}
\usepackage{xfrac}

\usepackage{fancyhdr}
\usepackage{enumerate}
\def\be{\begin{equation}}
\def\ee{\end{equation}}

\usepackage{tikz}
\usepackage{bbold}
\numberwithin{equation}{section}
\begin{document}
\selectlanguage{english}
\frenchspacing
\pagenumbering{roman}
\begin{center}
\null\vspace{\stretch{1}}

{ \Large {\bf 
Concerns about the replica wormhole\\ derivation of the island conjecture
}}
\\
\vspace{1cm}
Anna Karlsson
\vspace{1cm}

{\small
{\it
Institute for Theoretical Physics, Utrecht University,\\
Princetonplein 5, 3584 CC Utrecht, The Netherlands}}
\vspace{1.6cm}
\end{center}

\begin{abstract}
In a close look at the replica wormhole derivation of the island conjecture, we note a discrepancy in the derived matrix eigenvalues in the limit where the Hawking radiation entropy goes to zero. The small corrections to the density matrix for the radiation infer that ${\rm tr}(\rho_\textsf{R}^2)$, equally a sum over the eigenvalues squared, must be small ($\ll1$) at all times. This is in contradiction with an ensemble dominated by a density matrix with an eigenvalue close to unity.

Also, the radiation entropy has been approximated by the ensemble averaged entropy, as in a final state analysis. For $\rho_\textsf{R}=\rho_\textsf{R}(\sigma)$ with $\sigma$ the set of variables used in the average, the ensemble average is only an accurate estimate of the entropy after $\sigma$ has been measured. At the formation of each Hawking pair, there is a distribution over $\sigma$ that affects the entropy. Without a real time determination of $\sigma$, the radiation entropy follows Hawking's result during the evaporation. A question remains of how quickly such determinations can be assumed to occur. A complete argument for that the radiation entropy follows the Page curve should address this.
\end{abstract}
\vspace{\stretch{3}}
\thispagestyle{empty}

\newpage

\tableofcontents
\vspace{0.5cm}
\hrule
\pagenumbering{arabic}

\section{Introduction}
The recent derivation of the Page curve \cite{Page:1993wv,Page:2013dx} through the island conjecture \cite{Penington:2019npb,Almheiri:2019psf,Almheiri:2019hni} and the subsequent gravitational derivation of the same formula \cite{Penington:2019kki,Almheiri:2019qdq} have received much attention. Recent articles on the subject include \cite{Akers:2019nfi,Almheiri:2019yqk,Bousso:2019ykv,Almheiri:2019psy,Liu:2020gnp,Marolf:2020xie,Gautason:2020tmk,Anegawa:2020ezn,Sully:2020pza,Hartman:2020swn,Geng:2020qvw,Chen:2020uac,Bousso:2020kmy,Anous:2020lka,Engelhardt:2020qpv,Chen:2020tes,Hartman:2020khs,Liu:2020jsv,VanRaamsdonk:2020tlr,Akers:2020pmf,Stanford:2020wkf,Chen:2020hmv,Marolf:2020rpm,Goto:2020wnk,Colin-Ellerin:2020mva,Geng:2020fxl,Raju:2020smc}. The central property of these models --- the island conjecture and the gravitational path integral calculation using replica wormholes that gives the same formula for the Hawking radiation entropy --- is that they describe black hole evaporation as compatible with both unitarity and semiclassical physics at and outside the black hole event horizon. The small corrections theorem of \cite{Mathur:2009hf} is not sufficient to rule out unitarity of semiclassical black hole evaporation. Still, there are some question marks about whether or not the island conjecture indeed is compatible with semiclassical physics at the black hole event horizon. An important part of the consistency analysis of the suggested models is to directly connect the model properties to the conditions implied by semiclassical physics at the event horizon --- i.e. by that the Hawking pair production is nearly independent of the environment. In this text we connect the gravitational calculation of \cite{Penington:2019kki,Almheiri:2019qdq} to the conditions implied by nearly independent pair production on the density matrix of the radiation system, $\rho_\textsf{R}$. We address the question of whether or not the gravitational path integral calculation using replica wormholes is fully compatible with semiclassical physics at and outside the event horizon by showing that in the gravity theory, a black hole evaporation \emph{only} through nearly independent pair production cannot give a pure radiation system at late times. Consequently, the replica wormhole result of ${\rm tr}(\rho_\textsf{R}^n)=({\rm tr}(\rho_\textsf{R}))^n$ (at very late times for the evaporating black hole) is not a product of that type of evaporation, and requires some modified interpretation.

We identify two concerns that need to be addressed for a complete argument on the behaviour of the radiation entropy.
\begin{enumerate}
\item Fully independent Hawking pair production is represented by $\rho_\textsf{R}\propto \mathbb{1}$. Nearly independent pair production allows for small corrections to this. As detailed in \S\ref{s.small}, the eigenvalues $\lambda_{i,\textsf{R}}$ of the corrected $\rho_\textsf{R}$ are bounded by $\sum_i \lambda_{i,\textsf{R}}^2={\rm tr}(\rho_\textsf{R}^2)\ll1$ at late times. Meanwhile, a pure radiation system must have an eigenvalue $\sim 1$. Small corrections hence cannot accommodate a pure radiation system at (very) late times for the evaporating black hole, despite that the gravitational path integral calculation using replica wormholes gives that result.
\item At nearly independent pair production, the correlations between the two new modes (the interior and the radiation modes) and the environment cannot be deterministic. There has to be a probability distribution over different realisations of the correlations\footnote{In short, the second concern is the following (with the notation explained in \S \ref{s.setup}). At each pair formation, there must be a distribution over the value of each new, non-zero $R_{ij}$. Otherwise the pair formation is not nearly independent of the environment. By definition, the entropy of a distribution over $R_{ij}$ is different from the ensemble averaged entropy. For the radiation entropy to follow the Page curve, the system must evolve to reflect the required change in entropy. This evolution remains to be motivated in terms of what happens in the radiation system, which is a system that can be treated as separate from the black hole interior, and which is expected to be governed by semiclassical physics.}: $\rho_\textsf{R}=\int d\sigma \,p(\sigma)\rho_\textsf{R}(\sigma)$, where $\rho_\textsf{R}(\sigma)$ describes a specific realisation of the corrections, and $p(\sigma)<1$ is the probability for that realisation to occur. If there is no such distribution initially, at the pair production, then the pair production is not nearly independent of the environment in that it is either fully independent of or fully determined by the environment.

Any realisation of $\rho_\textsf{R}$ that can appear in the gravity theory is a $\rho_\textsf{R}(\sigma)$. This includes the different realisations of the ensemble discussed in \cite{Penington:2019kki}. To the purpose of our discussion, it does not matter if the ensemble average in the replica wormhole derivation of the Page curve is interpreted as due to that the gravity theory is dual to an ensemble of theories on the boundary, if the different realisations are interpreted as coming from a disorder average in a (self-averaging) single theory \cite{Pollack:2020gfa}, or due to a perturbative treatment of the gravitational theory. Any realisation of $\rho_\textsf{R}$ must be a $\rho_\textsf{R}(\sigma)$ with $p(\sigma)<1$ at the pair production, for the event horizon to be a region governed by semiclassical physics.

In addition, the (von Neumann) entropy of the radiation is always $S_{vN}(\rho_\textsf{R})$. This is equivalent to\footnote{We use same the notation for averaging as in \cite{Stanford:2020wkf}, for clarity. The average of a quantity $v$ can of course (for example) be written as either $\mathbb{E}[v]$, $\langle v\rangle$ or $\overline{v}$. Here, $\mathbb{E}[v(\sigma)]=\int d\sigma\, p(\sigma)v(\sigma)$.} $S_{vN}(\mathbb{E}[\rho_\textsf{R}(\sigma)])$. It is only \emph{after} a full determination of $\sigma$ (setting $p(\sigma)\in\{0,1\}$) that the ensemble averaged entropy $\mathbb{E}[S_{vN}(\rho_\textsf{R}(\sigma))]$ gives an accurate estimate of the radiation entropy (and then only if the deviations from the average are small). That the ensemble averaged entropy does not reflect the system entropy before a determination of $\sigma$ is a standard property. For completeness, we describe it in detail in \S\ref{s.ent}.

The replica wormhole setup in \cite{Penington:2019kki} has $\mathbb{E}[\rho_\textsf{R}(\sigma)]]\propto\mathbb{1}$. For the radiation entropy to follow the Page curve in that setting, {\it during the time of black hole evaporation} and in compatibility with semiclassical physics at the event horizon, $\sigma$ must be \emph{spontaneously} determined within the radiation system (i.e. within a semiclassical region) quickly enough and with a high enough accuracy for $S_{vN}(\rho_\textsf{R})$ to follow the Page curve instead of Hawking's calculation \cite{Hawking:1976ra}. A discussion regarding this is as of yet missing in the argumentation for that a gravitational calculation can reproduce the Page curve. Instead, the discussion has focussed on the ensemble averaged entropy, which corresponds to an analysis of a final state configuration.
\end{enumerate}

There are essentially three solutions to the first concern, regarding the evaporating black hole: approximate unitarity only (for any realisation of the radiation system), a (possibly small) remnant at the end of black hole evaporation, or new physics at the event horizon (e.g. at very late times). Either way, at least some replica wormhole configurations (topologies) are incompatible with semiclassical physics at and outside the event horizon, since they provide the result ${\rm tr}(\rho_\textsf{R}^n)=({\rm tr}(\rho_\textsf{R}))^n$ of \cite{Penington:2019kki,Almheiri:2019qdq}. Separate arguments for that some of the replica wormhole topologies should be excluded have been presented in \cite{Giddings:2020yes}. 

The most likely solution to the first concern is that the corrections are large at very late times for the evaporating black hole. In fact, that is even a property of the model of $\rho_\textsf{R}$ in \cite{Penington:2019kki}, where the corrections are of order $e^{-S/2}$ and hence $\mathcal{O}(1)$ when the entropy $S\rightarrow0$. However, this implies that the replica wormhole calculation gets out of control at those late times. So far, there has been no discussion on large corrections at the end of black hole evaporation in connection to the replica wormhole derivation of the island conjecture, or on how the replica wormhole topologies should be treated with respect to that.

There are four possible solutions to the second concern: a rapid, accurate determination of $\sigma$, new physics at the event horizon, non-unitarity, or an initial $\mathbb{E}[\rho_\textsf{R}(\sigma)]\not\propto \mathbb{1}$.

Regarding the second concern, we open up for a discussion on how the $\rho_\textsf{R}$ can be allowed to evolve during the evaporation. We present our thoughts on the matter in \S\ref{s.disc}. A discussion on (partial) determination of $\sigma$ already exists in\footnote{In refs. \cite{Marolf:2020xie,Marolf:2020rpm}, there is a discussion on how to (partially) fix their $\alpha$-states, where fixing $\alpha$ is related to fixing $\sigma$ in our notation, by using spacetime D-branes (similar to the D-branes and eigenbranes of \cite{Saad:2019lba,Blommaert:2019wfy}).} \cite{Marolf:2020xie,Marolf:2020rpm}. It is important to note that an analysis of the final state only (the state of $\rho_\textsf{R}$ a long time after evaporation, on $\mathcal{I}^+$) is insufficient, at least if a (partial or full) determination of $\sigma$ is allowed prior to that the final state forms. Unitarity must hold at all times, including during the process of evaporation.

Note that even if gravity computes the average of the entropy --- which seems very likely based on that it gives the Page curve result --- it is still important to understand the physics that is required to get to that result. There is a difference between the ensemble average of the entropy (the average of $S(\rho_\textsf{R}(\sigma))$ with respect to $\sigma$) and the entropy of the distribution of the possible final states, $S(\rho_\textsf{R})$. At nearly independent pair formation, there is a distribution over different corrections in $\rho_\textsf{R}$, and that distribution affects the entropy of the radiation; the system that the Page curve was derived for in \cite{Page:1993wv}. A good understanding of how $S(\rho_\textsf{R})$ evolves to the ensemble averaged entropy is required for a proper understanding of the result of the replica wormhole calculation.

To illustrate our two concerns about the recent derivations of the Page curve, we begin with a brief summary of the replica wormhole setup of \cite{Penington:2019kki} in \S\ref{s.setup}, since that is the most convenient framework among the existing ones \cite{Penington:2019kki,Almheiri:2019qdq,Marolf:2020xie} for direct analyses of properties of density matrices with small corrections. We then present our fist concern in detail in \S\ref{s.small} and the second one in \S\ref{s.ent}. Finally, we discuss the implications of these concerns and possible solutions in \S\ref{s.disc}.

\section{The replica wormhole setup}\label{s.setup}
The replica wormhole setup is as follows. The full system is in a pure state
\begin{subequations}\label{eq.setup}
\be
|\Psi\rangle=\frac{1}{\sqrt{k}}\sum_{i=1}^k|\psi_i\rangle_\textsf{B}| i\rangle_\textsf{R}
\ee
after $k$ Hawking modes have been emitted from the black hole. The states $|i\rangle$ are part of the Hawking radiation system $\textsf{R}$ and form an orthonormal basis. The $|\psi_i\rangle$ is the black hole state connected to $|i\rangle$ through the Hawking pair production. The density matrix for the radiation is
\be
\rho_\textsf{R}={\rm tr}_\textsf{B}(|\Psi\rangle\langle\Psi|)= \frac{1}{k} \sum_{i,j=1}^k \langle\psi _i|\psi_j\rangle_\textsf{B}\,|j\rangle\langle i|_\textsf{R}\,,
\ee 
a Hermitian, positive semi-definite matrix of trace one. The black hole interior states are set to be almost, but not quite, orthogonal,
\be
\langle\psi _i|\psi_j\rangle=\delta_{ij}+x R_{ij}\,,\qquad |xR_{ij}|\ll1\,,\quad \mathbb{E}\left[R_{ij}\right]=0\,,\quad\mathbb{E}\left[|R_{ij}|^2\right]=1\,,
\ee
\end{subequations}
where $x$ is real and the toy model in \cite{Penington:2019kki} has $x=e^{-S/2}$. The terms $R_{ij}$ represent random variables, and the $xR_{ij}$ in $\langle\psi _i|\psi_j\rangle$ represents a noise term, providing (exponentially) small corrections \cite{Penington:2019kki,Almheiri:2019qdq} to the state overlap. The noise terms are introduced to accommodate the gravitational path integral results of
\be\label{eq.cond}
\langle\psi _i|\psi_j\rangle=\delta_{ij}\,,\qquad |\langle\psi _i|\psi_j\rangle|^2=\delta_{ij} + x^2\,,\qquad x\neq0\,,
\ee
which are consistent when the path integral is interpreted to give disorder averaged results, i.e. $\mathbb{E}[\langle\psi _i|\psi_j\rangle]$ and $\mathbb{E}[|\langle\psi _i|\psi_j\rangle|^2]$ instead of $\langle\psi _i|\psi_j\rangle$ and $|\langle\psi _i|\psi_j\rangle|^2$ in \eqref{eq.cond}. The average corresponds to an ensemble average over different density matrices $\rho_\textsf{R}$. Here, the first condition in \eqref{eq.cond} follows from a gravitational computation of the single overlap, $\langle\psi _i|\psi_j\rangle$. The second condition, or more precisely the fluctuating term $R_{ij}$, is allowed by the term $|\langle\psi _i|\psi_j\rangle|^2$, and is what enables a reproduction of the Page curve; a result that is provided (separately) by the gravitational path integral calculation over different replica topologies. \cite{Penington:2019kki}

The replica wormhole setup as outlined right above is the one given in \cite{Penington:2019kki}, and constitutes a description of how to think of the states in the gravity theory. However, that description in itself is not necessary to obtain the result of the Page curve. It is a setup modelled on the gravitational path integral results. For example, the values of $xR_{ij}$ are never identified; their presence is inferred from \eqref{eq.cond}. The actual computation, provided in \cite{Penington:2019kki,Almheiri:2019qdq}, is a gravitational path integral over different topologies allowed between replicas of the system $\rho_\textsf{R}$.

The replica trick for density matrices uses the fact that the von Neumann entropy $S_{vN}$ can be obtained from the R\'enyi $n$-entropy $S_n$ through
\be
S_{vN}(\rho)=\lim_{n\rightarrow1}S_n(\rho)\,,\qquad S_n(\rho)=\frac{1}{1-n}\log {\rm tr}(\rho^n)\,.
\ee
The number of replicas is $n$, and the initial calculation of integer $n$ is analytically continued to non-integer $n$ to obtain the limit. Descriptions of how the gravitational path integral calculation is performed can be found in \cite{Penington:2019kki,Almheiri:2019qdq}. For the analysis below, the present level of detail is sufficient.

\section{Small corrections vs a pure radiation system}\label{s.small}
The Page curve for an evaporating black hole is quite different from that of an eternal black hole. A key characteristic of the former scenario is that the radiation entropy goes to zero at very late times. The purity of the final configuration is reproduced by the gravitational path integral calculation using replica wormholes in that the calculation gives the result
\be
{\rm tr}(\rho^n_\textsf{R})=({\rm tr}(\rho_\textsf{R}))^n=1
\ee
for certain wormhole topologies \cite{Penington:2019kki,Almheiri:2019qdq}. This result is to be interpreted as produced by an ensemble average. In some discussions on replica wormhole setups, e.g. \cite{Almheiri:2020cfm,Marolf:2020rpm}, the discussed $\rho_\textsf{R}$ is not normalised to have trace one. In that case, purity is given by the first equality above. However, a completely general setup of small corrections to nearly independent Hawking pair production is represented by \eqref{eq.setup}, with ${\rm tr}(\rho_\textsf{R})=1$ and without the restrictions on $\mathbb{E}\left[R_{ij}\right]$ and $\mathbb{E}\left[|R_{ij}|^2\right]$. This is the setup we treat in this section, without loss of generality.

Note the following. The fact that (for positive integers $n$) 
\be
{\rm tr}(\rho^n)=\sum_i\lambda_i^n
\ee
where $\lambda_i$ are the eigenvalues of $\rho$ (with $\lambda_i\geq0$) means that
\be\label{eq.relation}
{\rm tr}(\rho^2_\textsf{R})=\frac{1}{k^2}\sum_{ij}|\langle\psi_i|\psi_j\rangle|^2=\sum_i\lambda_{i,\textsf{R}}^2= k^{-1}+\frac{1}{k^2}\sum_{ij}|xR_{ij}|^2
\ee
for every realisation of the $\rho_\textsf{R}$. To get the expression to the far right, we have used that ${\rm tr}(\rho_\textsf{R})=1$ sets $\sum_i R_{ii}=0$. For large $k$, \eqref{eq.relation} sets an upper bound on each eigenvalue,\footnote{This upper bound is more precisely given by a mean value: $\lambda_{i,\textsf{R}}^2\lesssim \sum_{ij}|xR_{ij}|^2/k^2$, and rigorously by \eqref{eq.relation}.}
\be\label{eq.cond1}
\lambda_{i,\textsf{R}}\,: \quad \lambda_{i,\textsf{R}} \lesssim \max\{|xR_{ij}|\}\quad\text{when}\qquad k^{-1}\ll \sum_{ij}|xR_{ij}|^2/k^2\,.
\ee
Meanwhile, for the ensemble result to be pure, the ensemble must be dominated by at least one $\rho_\textsf{R}$ with an eigenvalue close to unity. We therefore have two conflicting requirements,
\be\label{eq.cont}
{\rm tr} (\rho_\textsf{R}^n)\sim1\,\Rightarrow \,\exists\, \lambda_{i,\textsf{R}}\sim 1\qquad \text{vs}\qquad\lambda_{i,\textsf{R}} \lesssim \max\{|xR_{ij}|\}\ll1\,,
\ee
at late times, when $k^{-1}$ obeys \eqref{eq.cond1}. To the left is the requirement for a nearly pure radiation system. To the right is the requirement that the eigenvalues of each realisation of the $\rho_\textsf{R}$ in \eqref{eq.setup} are restricted to be small {\it in the same sense} that the corrections to $\langle\psi_i|\psi_j\rangle=\delta_{ij}$ are restricted to be small.

Equation \eqref{eq.cont} illustrates a tension between an ensemble averaged entropy $\mathbb{E}[S_{vN}(\rho_\textsf{R})]$ following the Page curve of an evaporating black hole, with $\mathbb{E}[S_{vN}(\rho_\textsf{R})]\rightarrow0$ at very late times, and the eigenvalues that individual realisations of $\rho_\textsf{R}$ are allowed to have under the condition of small corrections to the Hawking pair production. $|xR_{ij}|\ll1$ is required for the Hawking pair production to be nearly independent, i.e. for it to represent pair production in a semiclassical region.

By \eqref{eq.cond1}, a restriction to small corrections is equivalent to $S_{vN}(\rho_\textsf{R})>0$ for all realisations of $\rho_\textsf{R}$ at late times, and thus also for the ensemble averaged entropy. The radiation entropy can be very small, but the system can never be pure. In that setting, unitarity is at best approximate, never exact --- not only with respect to individual realisations of the radiation system, which have been allowed to fluctuate around the average, but for any average as well.

\section{Radiation entropy vs the final state entropy}\label{s.ent}
The result for the radiation entropy given by the gravitational path integral calculation using replica wormholes is interpreted as an ensemble average
\be\label{eq.estS}
\mathbb{E}\left[S_{vN}(\rho_\textsf{R})\right]\,.
\ee
This is an \emph{estimate} of the system entropy, after the system has ended up in one of several possible configurations. In the gravitational path integral calculation using replica wormholes, this entropy follows the Page curve. The deviation from the average entropy by any given realisation of the ensemble is expected to be small \cite{Penington:2019kki} or exponentially small \cite{Marolf:2020rpm}, and the existence of such single realisations is considered to be a guiding principle \cite{Penington:2019kki} in the conjecture that there should exist non-averaged theories where the radiation entropy follows the Page curve \cite{Penington:2019kki,Almheiri:2019qdq}.

Now, consider the following for a black hole. A key property of nearly independent Hawking pair production is that upon formation of a new Hawking pair, the magnitudes of the correlations between the new modes (radiation mode and interior mode) and the other modes in the environment can take a range of values. In \eqref{eq.setup}, this is inferred from the presence of a distribution over $R_{ij}$ in $\langle \psi_i|\psi_j\rangle$. If the corrections to $\langle \psi_i|\psi_j\rangle=\delta_{ij}$ are uniquely determined, the pair production is not nearly independent --- it is either fully independent of the environment (no corrections), or fully determined by the environment. In the presence of an ensemble average, there are different configurations that the radiation system can take; the ensemble average is over those configurations.

An important assumption is made when one considers the ensemble averaged entropy of a system. While the ensemble averaged entropy is an estimate of the system entropy, it is only accurate {\it after the configuration of the system has been determined.} To be precise, the variables that are averaged over in the ensemble average must have been measured. A determination through interactions within the system is sufficient.

For a simple illustration of this, consider the following classical example. Typically, when a state is formed that can take two sets of values $i$ and $\sigma$, the Shannon entropy of that state is
\begin{subequations}
\be\label{eq.Sh}
S_{Shannon}=-\sum_{i,\sigma} p(i,\sigma)\log p(i,\sigma)\,,
\ee
where $p(i,\sigma)$ is the probability for each of the possible, different configurations. For example, consider a scenario where one has randomly thrown either a coin or a die, and denote the outcome (numbers 1 through 6, heads or tails) by $i$ and the shape (coin or die) by $\sigma$. 
It is only \emph{after} the state of $\sigma$ has been determined that one arrives at the configuration for which the entropy can be estimated by the ensemble average over $\sigma$,
\be\label{eq.cest}
\mathbb{E}\left[-\sum_i p(i|\sigma)\log p(i|\sigma)\right]\,,\qquad p(i,\sigma)=p(i|\sigma)p(\sigma)\,,
\ee
\end{subequations}
where $p(i|\sigma)$ is the conditional probability of outcome $i$ given the outcome $\sigma$. To compare with the calculation of the radiation entropy, denote the different configurations of the radiation system by $\rho_\textsf{R}(\sigma)$, where $\sigma$ is the set of correlations: $\sigma=\{xR_{ij}\}$. In looking at the ensemble averaged entropy of the radiation system, \eqref{eq.estS}, one considers the scenario \eqref{eq.cest} instead of \eqref{eq.Sh}, and there is an assumption of that the $\sigma=\{xR_{ij}\}$ have been determined.

To be precise, in considering the ensemble averaged entropy one assumes that the ensemble part of the configuration in question is discernible (determined, measured), e.g. in a final state configuration, so that no probability distribution over the various different configurations of the ensemble feeds into the entropy, giving additional degrees of freedom to account for in the entropy.

In terms of the Hawking radiation entropy, one does not merely look at a system that has ended up in a final configuration. The Hawking pair production takes place during the process of black hole evaporation, i.e. during the time span when the entropy $S_{vN}(t)$ of the radiation system needs to follow the Page curve for unitarity to hold. At the pair production itself, there is a distribution over the corrections $\{xR_{ij}\}$. At that point, and up until the configuration of $\{xR_{ij}\}$ in \eqref{eq.setup} has been measured, the probability distribution over the various configurations feeds into the radiation entropy, making it larger than the estimate of the entropy given by the ensemble average --- but by how much?

When correcting for the additional degrees of freedom in an estimate of the radiation entropy, it is important to recall that the entropy is only captured by \eqref{eq.Sh} when the states described by $(i,\sigma)$ are orthogonal. For this reason, is is incorrect to e.g. base the entropy on the eigenvalues $\lambda_i(\sigma)$ of each matrix $\rho_\textsf{R}(\sigma)$ through $p(i,\sigma)=\lambda_i(\sigma)p(\sigma)$. In density matrix theory the eigenstates of the matrix (for a division into $\rho_\textsf{R}(\sigma)$, matrices) must be orthogonal. As $\rho_\textsf{R}$ is built from the radiation states $|i\rangle$, there are only $k$ orthogonal eigenvectors for all of the $\rho_\textsf{R}(\sigma)$. Consequently, the system is described by a single matrix, $\mathbb{E}[\rho_\textsf{R}(\sigma)]$.

One can identify upper and lower bounds for the deviation from the ensemble averaged entropy as follows. An upper bound on the entropy contribution from the probability distribution over the different configurations is (as follows from the argument right above) that the radiation entropy follows Hawking's calculation instead of the Page curve. That case corresponds to the scenario when the correlation configurations provided by $\langle\psi_i|\psi_j\rangle$ are not determined at all, at any point in time, and $\mathbb{E}[\rho_\textsf{R}(\sigma)]\propto\mathbb{1}$. A lower bound is that each such configuration is fully determined between Hawking particle emissions. In that case, the effect would be very small. The corrections in \eqref{eq.setup} only affect the entropy at large $k$ ($k\gtrsim x^{-2}$). The addition of a new mode $(k-1)\rightarrow k$ in \eqref{eq.setup} would correspond to $\langle\psi_i|\psi_k\rangle=\delta_{ik}$ and an added eigenvalue of $1/k$ in combination with a $(1-1/k)$ rescaling of the previous eigenvalues:
\begin{gather}\begin{aligned}
-S_{vN}(\rho_\textsf{R})&=\sum_{i=1}^{k-1}\lambda_{i,\textsf{R}}\log\lambda_{i,\textsf{R}}\\
&\rightarrow \left(1-1/k\right)\sum_{i=1}^{k-1}\lambda_{i,\textsf{R}}\log\lambda_{i,\textsf{R}}+(1-1/k)\log(1-1/k)+(1/k)\log(1/k)\,,
\end{aligned}\end{gather}
with a very small deviation for large $k$.

Despite that a best case scenario is quite faithful to the ensemble averaged result for the radiation entropy, it is necessary to consider what it would require of the physical system. The measurements required for determining $\{xR_{ij}\}$ can occur through interactions within the system (as opposed to by an `external' observer) but must take place in a semiclassical region for new physics (and other explanation models, besides the island conjecture) not to come into play. Moreover, the correlations $\{\delta_{ij}+xR_{ij}\}$ need to be determined with high precision. The corrections are (exponentially) small, and an uncertainty in the determined $xR_{ij}$ reasonably propagates into the entropy with a similar impact as the actual, exact corrections. Of course, a partial determination (a narrowing down of the range of the actual values) of $\{xR_{ij}\}$ could reduce the entropy, but a very precise determination would be required to reproduce the Page curve.

\section{Discussion}\label{s.disc}
We have highlighted two concerns regarding derivations of the Page curve that keep semiclassical physics at and outside the black hole event horizon, by focussing on explicit realisations of density matrices for the Hawking radiation system, and on properties they must obey due to nearly independent pair production at the event horizon. These concerns are valid in general, not only for the island conjecture and the gravitational path integral calculation using replica wormholes. First, small corrections to $\rho_\textsf{R}\propto\mathbb{1}$ set $S_{vN}(\rho_\textsf{R})>0$ through an upper bound on the eigenvalues. Hence, the Page curve for the evaporating black hole at late times cannot be reproduced exactly. Second, in analyses of the radiation entropy it is insufficient to consider only the final state of $\rho_\textsf{R}$, as unitarity must hold at all times. As the radiation forms, semiclassicality requires $\rho_\textsf{R}=\int d\sigma\, p(\sigma)\rho_\textsf{R}(\sigma)$ with $p(\sigma)<1$, where each $\rho_\textsf{R}(\sigma)$ is a realisation of the radiation system with all the corrections $xR_{ij}$ uniquely specified. It is this $S_{vN}(\rho_\textsf{R})$ that must follow the Page curve, not only a final state with $p(\sigma)\in\{0,1\}$. A suitable evolution of $S_{vN}(\rho_\textsf{R})$ remains to be motivated.

As observed in \eqref{eq.cond1}, the eigenvalues of the setup in \eqref{eq.setup} (without a restriction to specific averages for $R_{ij}$) have an upper bound. In specific, at late times ${\rm max}\{\lambda_{i,\textsf{R}}\}$ is restricted to be small (at least) in the same sense that the corrections $|x R_{ij}|$ are restricted to be small. Hence $S_{vN}(\rho_\textsf{R})>0$. The upper bound on the eigenvalues and the ansatz of \eqref{eq.setup} hold as long as the black hole evaporation takes place through Hawking pair production that is nearly independent of the environment. A discussion with respect to the property $S_{vN}(\rho_\textsf{R})>0$, effectively the contradiction of \eqref{eq.cont}, is so far lacking in the articles on the derivations (holographic and gravitational) of the Page curve, where the exact Page curve for the evaporating black hole is derived, including $S_{vN}(\rho_\textsf{R})=0$. In the gravitational calculation, the purity of the radiation system is captured by the result of ${\rm tr}(\rho_\textsf{R}^n)=({\rm tr}(\rho_\textsf{R}))^n$. Based on our eigenvalue argument alone, it is clear that some of the replica wormhole topologies include deviations from semiclassical physics at or outside the black hole event horizon, despite that the gravitational path integral calculation using replica wormholes initially does not appear to include such deviations.

Note that the inclusion of all replica wormhole topologies in \cite{Penington:2019kki,Almheiri:2019qdq} constitutes a \emph{model choice}. The gravitational path integral calculation and the replica trick are standard calculations, but the replica wormhole topologies constitute an introduced element --- a priori assumed to be compatible with semiclassical physics at and outside the event horizon, but not restricted to be so. A key question here is if the (full) gravitational path integral calculation using replica wormholes can be trusted for $S_{vN}(\rho_\textsf{R})\geq\epsilon$, for some $\epsilon>0$ (e.g. until the black hole is of Planckian size), or if some wormhole topologies need to be excluded altogether for compatibility with semiclassical physics at and outside the event horizon. In the former scenario, one could perhaps settle for approximate unitarity instead of exact, or some remnant at the end of black hole evaporation. Already, the individual realisations of $\rho_\textsf{R}$ that are part of the ensemble discussed in \cite{Penington:2019kki} describe fluctuations away from the Page curve. If some wormhole topologies need to be excluded, a modified analysis of the gravitational derivation of the Page curve would be required, if that can reproduce the Page curve. In extension, this is also relevant for the holographic formula. The island conjecture is a novel take on the quantum extremal surface prescription \cite{Engelhardt:2014gca}, a generalisation of the Ryu--Takayanagi formula \cite{Ryu:2006bv,Hubeny:2007xt}, which in turn can be derived through a gravitational path integral calculation \cite{Lewkowycz:2013nqa,Faulkner:2013ana,Dong:2016hjy,Dong:2017xht}. The same is expected to hold for the island conjecture. If the island conjecture cannot be derived in that way, with the region at and outside the event horizon governed by semiclassical physics, there would be a question of if a disconnected quantum extremal surface (as introduced in the island conjecture) fits with semiclassical physics in that same region.

The concern regarding the distribution over the small corrections during black hole evaporation is of a more evasive nature than the upper bound on the eigenvalues, but it is still highly relevant. There is a marked difference between an analysis of the final state and of the $\rho_\textsf{R}$ present during the time of black hole evaporation --- and unitarity ought to hold at all times, not only for an analysis of what the final state says about the evaporation process. This concern also has more far-reaching consequences, as it applies to the Page curve for both the evaporating black hole and the eternal black hole, and since a determination of $\sigma=\{xR_{ij}\}$ is crucial for $S_{vN}(\rho_\textsf{R})$ to follow anything like the Page curve when $\mathbb{E}[\rho_\textsf{R}(\sigma)]\propto\mathbb{1}$, as is the case for the derivation in \cite{Penington:2019kki}. A central observation is also that a distribution over the corrections is implied by a presence of nearly independent pair production. It is not possible to evade a determination of the set $\{xR_{ij}\}$ by considering a single theory, e.g. some higher-order completion of the ensemble average, since any single theory specifying $p(\sigma)\in\{0,1\}$ immediately at the pair production would be incompatible with an event horizon region governed by semiclassical physics.

We would welcome a discussion in the science community about whether or not the corrections $\{xR_{ij}\}$ can be assumed to be determined in such a way that $S_{vN}(\rho_\textsf{R})$ follows the Page curve during the process of black hole evaporation. Key issues here include that the determination of $\sigma$ must take place in a region of semiclassical physics, occur spontaneously, at a rate and with an accuracy high enough to reproduce the Page curve. Keep in mind that the determination of $\sigma$ is a reduction of $\rho_\textsf{R}$ and a determination of the correlations between the Hawking modes, equivalently a determination of the specific small corrections to the density matrix of the radiation system. The determination of the $\{xR_{ij}\}$ can be expected to be made difficult both by the smallness of the corrections, and by that the entity that must be materially altered is $\mathbb{E}[\rho_\textsf{R}(\sigma)]$. Considering this, it would seem unlikely that the radiation entropy would follow the Page curve during the process of black hole evaporation.

However, it is not only the speed and precision required that is of concern. When the initial pair production (disregarding a possible, subsequent determination of $\sigma$) gives $\rho_\textsf{R}\propto\mathbb{1}$, the determination is equivalent to a process that takes uncorrelated modes (radiation modes, as described by $\rho_\textsf{R}$) to entangled ones. In semiclassical physics, a state must first disentangle from all other modes before entangling with a new mode, by monogamy of entanglement, but the required determination of $\sigma$ would have to include an (effective) entangling of the Hawking modes without disturbing the state of each mode. Each radiation mode would have to remain fully entangled with its interior partner, throughout the process. Semiclassically, this is counterintuitive, as noted in \cite{Karlsson:2020uga}. Suggested scenarios to get around that type of contradiction include \cite{Raju:2016vsu,Raju:2018zpn}.

Another option could be an initial $\mathbb{E}[\rho_\textsf{R}(\sigma)]$ (again, prior to any determination of $\sigma$) that follows the Page curve. That scenario fits within what is allowed by small corrections (up to the first concern detailed above) but would likely be incompatible with nearly independent pair production, since that implies a \emph{weak} influence of the environment on the pair production. Small corrections is one requirement. Reasonably, a non-deterministic sign of each correction is also required, so that $\mathbb{E}[\rho_\textsf{R}(\sigma)]\propto\mathbb{1}$, initially.
\\\\
We argue that the holographic and gravitational derivations of the Page curve need to be expanded to address the concerns we have described above, in order for the derivations to be be compatible with semiclassical physics at and outside the black hole event horizon. Central questions include the following: Which replica wormhole topologies are allowed? How to reconcile the late time property $S_{vN}(\rho_\textsf{R})>0$, inferred by the small corrections, with unitarity for the evaporating black hole? Apart from what the ensemble average of the final states specifies, can the small corrections in $\rho_\textsf{R}$ be determined quickly enough and with high enough accuracy, within the radiation system, to accommodate a radiation entropy that approximately follows the Page curve \emph{during} the process of evaporation, i.e. without significant deviations from the Page curve?

\section*{Acknowledgements}
I thank D. Stanford and S. Vandoren for useful comments on a draft of this paper.

\providecommand{\href}[2]{#2}\begingroup\raggedright\endgroup

\end{document}